\documentclass[aps,pra,superscriptaddress,floatfix,twocolumn]{revtex4-1}
\usepackage{enumerate}
\usepackage{mathtools}
\usepackage{amsmath}
\usepackage{graphicx}
\usepackage{epsfig}
\usepackage{sidecap}
\usepackage{hyperref}
\usepackage{color}
\usepackage{subfigure,array}
\usepackage{verbatim}
\usepackage{multirow}
\usepackage{lipsum}
\usepackage{bm}
\usepackage{float}

\begin{document}

\title{Interaction of One-Dimensional Quantum Droplets with Potential Wells
and Barriers}
\author{Argha Debnath}
\affiliation{Department of Physics, School of Engineering and Applied Sciences, Bennett
University, Greater Noida, UP-201310, India}
\author{Ayan Khan}\thanks{ayan.khan@bennett.edu.in}
\affiliation{Department of Physics, School of Engineering and Applied Sciences, Bennett
University, Greater Noida, UP-201310, India}
\author{Boris Malomed}
\affiliation{Department of Physical Electronics, School of Electrical Engineering,
Faculty of Engineering, Tel Aviv University, Ramat Aviv 69978, Israel}\affiliation{
Instituto de Alta Investigaci\'{o}n, Universidad de Tarapac\'{a}, Casilla
7D, Arica, Chile}

\begin{abstract}
We address static and dynamical properties of one-dimensional (1D) quantum
droplets (QDs) under the action of local potentials in the form of narrow
wells and barriers. The dynamics of QDs is governed by the 1D
Gross-Pitaevskii equation including the mean-field cubic repulsive term and
the beyond-mean-field attractive quadratic one. In the case of the well
represented by the delta-function potential, three exact stable solutions
are found for localized states pinned to the well. The Thomas-Fermi
approximation for the well and the adiabatic approximation for the collision
of the QD with the barrier are developed too. Collisions of incident QDs
with the wells and barriers are analyzed in detail by means of systematic
simulations. Outcomes, such as fission of the moving QD into transmitted,
reflected, and trapped fragments, are identified in relevant parameter
planes. In particular, a counter-intuitive effect of partial or full rebound
of the incident QD from the potential well is studied in detail and
qualitatively explained.
\end{abstract}

\maketitle

\section{Introduction}
Quantum droplets (QDs) in binary Bose-Einstein condensates (BECs) originate
from the interplay of the mean-field (MF) interactions and Lee-Huang-Yang
(LHY) corrections to them, induced by quantum fluctuations \cite{LHY}. As
first demonstrated by Petrov \cite{petrov1}, this setting is modeled by the
modified Gross-Pitaevskii equation (GPE) which includes the MF
self-attractive cubic term and the repulsive LHY quartic one. This
prediction was followed by the creation of QDs in binary homoatomic \cite%
{Cabrera2018,Leticia,Inguscio,ferioli} and heteroatomic \cite{Chiara} BECs,
as well as in dipolar condensates \cite{Schmitt2016,Chomaz2016}. Actually,
QDs represent a new quantum state of matter, see reviews \cite%
{Sandy,debnath5}.

The dimensional reduction 3D $\rightarrow $ 2D and 3D $\rightarrow $ 1D for
BEC under the action of a tight confining potential essentially changes the
form of the effective GPE \cite{Petrov2016,Ilg,PRA103_013312,edmonds}. In
contrast with the 3D setting, in the 1D limit the LHY term has the
self-attraction sign, and is quadratic, rather than quartic \cite{Petrov2016}%
. However, it is also possible to extend the 3D framework to a
quasi-one-dimensional (Q1D) geometry by tuning the transverse confinement
suitably. In this situation, the repulsive quartic LHY term prevails to
support the droplet formation in both homogeneous and inhomogeneous systems
\cite{debnath1,debnath4}. Similarly, the quartic term provides
stabilization of Q2D bound states pulled to the center by the singular
potential $\sim -1/r^{2}$ \cite{Viskol}.

On the contrary, in the proper 1D system the residual self-repulsive MF term
competes with the quadratic self-attraction, making it possible to create
broad QD states with the equilibrium density in their inner quasi-flat parts
\cite{malomed,Tylutki}. The fact that the density cannot exceed the
equilibrium value implies that the quantum matter filling the QDs is an
incompressible liquid, therefore localized states are named
\textquotedblleft droplets".

The next natural step in the studies of the droplet dynamics is to
investigate collisions between moving QDs, which was realized experimentally
(in the 3D geometry) in Ref. \cite{ferioli}. In the 1D setting, simulations
of the corresponding GPE with the cubic-quadratic nonlinearity demonstrate
that the collision between slowly moving droplets with zero phase difference
between them leads to their merger into a single one, in an excited
(breathing) state, while the collision between QDs with phase difference $%
\pi $ ends up with quasi-elastic rebound \cite{malomed}. Collisions at large
velocities tend to produce quasi-elastic outcomes, i.e., passage of the QDs
through each other, similar to collisions between usual matter-wave solitons
\cite{Cornish}.

Another physically relevant problem is the consideration of collisions of an
incident QD with a confined attractive or repulsive defect, which may be
represented by a localized potential well or barrier, respectively. In the
experiment, such defects can be created, respectively, by narrow red- or
blue-detuned laser beams illuminating the condensate \cite%
{Cornish-experiment}. For moving solitons this problem was considered, in
various forms, in the framework of the nonlinear Schr\"{o}dinger equation
(NLSE) \cite{sakaguchi,goodman,lee,brand,Carr,ching,Harel,Khawaja}. In BEC,
collisions of matter-wave solitons with barriers were studied both
theoretically \cite{Helm,Khayk,two-comp,Callum} and experimentally \cite%
{Cornish-experiment}. In particular, much interest was drawn to the use of
narrow potential barriers as splitters in the design of soliton
interferometers \cite{Martin,Ahufinger,Helm2,Cornish2,Gardiner,HS}. Bound
states of low-dimensional QDs interacting with defects were recently
addressed in Refs. \cite{abdullaev,Macri,sayak}. Removable potential
barriers can be used for preparation of inputs initiating collisions between
condensates \cite{pylak}.

In this work, we aim to study interactions of QDs with potential wells and
barriers in the framework of the one-dimensional LHY-corrected GPE with the
cubic-quadratic nonlinearity. The corresponding model is formulated in Sec. %
\ref{model}. Analytical results obtained for the model are collected in Sec. %
\ref{analytics}. In particular, we consider quiescent QDs trapped in
potential wells. For the narrow well represented by the delta-function
potential we present three exact stable solutions for localized states
pinned to the well. The Thomas-Fermi and adiabatic approximations are
considered too, for pinned and moving QDs, respectively. Collisions of
moving QDs with the barrier or well are studied by means of systematic
simulations in Sec. \ref{numerics}. The outcome of the collisions is
quantified by computing shares of the initial norm in the transmitted,
reflected and trapped waves (trapping takes place in the case of the
potential well). We also plot maps, in the plane of the norm and velocity of
the incident QD, for values of the maximum strength of the potential well
admitting full transmission, and minimum strength required for the
observation of a counter-intuitive effect, in the form of the \textit{%
complete reflection} of the moving QD from the attractive potential. The
paper is concluded by Sec. \ref{conclusion}.

\section{The model}

\label{model}

In this section we summarize the 1D theoretical model. We start from the
binary BEC with equal self-repulsion coefficients in both components, $%
g_{11}=g_{22}=g>0$, and equal numbers of atoms in them. This setting makes
it possible to consider the symmetric configuration, with equal MF\ wave
functions of the components, $\psi _{1}=\psi _{2}\equiv \psi $. Then,
defining a small difference between the inter-component attraction strength (%
$g_{12}<0$) and intra-component self-repulsion, $\delta g=g+g_{12}>0$, with $%
\delta g\ll g$, one can derive the effective 1D GPE including the beyond-MF
correction (the quadratic term) as \cite{Petrov2016}:

\begin{equation}
i\hbar \frac{\partial \psi }{\partial t}=-\frac{\hbar ^{2}}{2m}\frac{%
\partial ^{2}\psi }{\partial x^{2}}+f(x)\psi +\delta g|\psi |^{2}\psi -\frac{%
\sqrt{2m}}{\pi \hbar }g^{3/2}|\psi |\psi ,  \label{1dbgp}
\end{equation}%
where $f(x)$ represents an external potential, and $m$ is the atomic mass.

Equation (\ref{1dbgp}) determines characteristic units of length, $x_{0}$,
time, $t_{0}$, energy, $E_{0}$, and wave function,

\begin{eqnarray}
&&x_{0}=\frac{\pi \hbar ^{2}\sqrt{\delta g}}{\sqrt{2}mg^{3/2}},t_{0}=\frac{\pi
^{2}\hbar ^{3}\delta g}{2mg^{3}},\nonumber\\&&E_{0}=\frac{\hbar ^{2}}{mx_{0}^{2}}=\frac{%
\hbar }{t_{0}},\psi _{0}=\frac{\sqrt{2m}}{\pi \hbar \delta g}g^{3/2},
\label{units}
\end{eqnarray}%
which suggests rescaling $t\equiv t_{0}\tilde{t}$, $x\equiv x_{0}\tilde{x}$,
$\psi \equiv \psi _{0}\tilde{\psi}$, $f^{\prime }(x)\equiv \tilde{f}(\tilde{x%
})/E_{0}$. In this notation, Eq. (\ref{1dbgp}) is cast in the normalized
form (where the tildes are omitted):

\begin{equation}
i\frac{\partial \psi }{\partial t}=-\frac{1}{2}\frac{\partial ^{2}\psi }{%
\partial x^{2}}+f(x)\psi +|\psi |^{2}\psi -|\psi |\psi ,  \label{dbgp}
\end{equation}%
with Hamiltonian (energy),%
\begin{eqnarray}
&&E=\int_{-\infty }^{+\infty }\left[ \frac{1}{2}\left\vert \frac{\partial \psi
}{\partial x}\right\vert ^{2}+f(x)|\psi (x)|^{2}-\frac{2}{3}|\psi (x)|^{3}+%
\frac{1}{2}|\psi (x)|^{4}\right] dx\nonumber\\&&\equiv E_{\text{\textrm{kin}}}+E_{\text{%
\textrm{int}}}, \label{H}
\end{eqnarray}%
which includes the kinetic, alias gradient (first), and interaction
(potential) terms.

Stationary solutions for QDs with chemical potential $\mu <0$, produced by
Eq. (\ref{dbgp}), are looked for in the usual form,%
\begin{equation}
\psi \left( x,t\right) =\exp \left( -i\mu t\right) \phi (x)\,,  \label{phi}
\end{equation}%
where real stationary wave function $\phi (x)>0$ satisfies the equation%
\begin{equation}
\mu \phi =-\frac{1}{2}\frac{d^{2}\phi }{dx^{2}}+f(x)\phi +\phi ^{3}-\phi
^{2}.  \label{stationary}
\end{equation}%
In particular, the free-space version of Eq. (\ref{stationary}), with $%
f(x)=0 $, gives rise to the known family of exact QD solutions \cite%
{Petrov2016},
\begin{figure}[tbp]
\begin{center}
\includegraphics[scale=0.25]{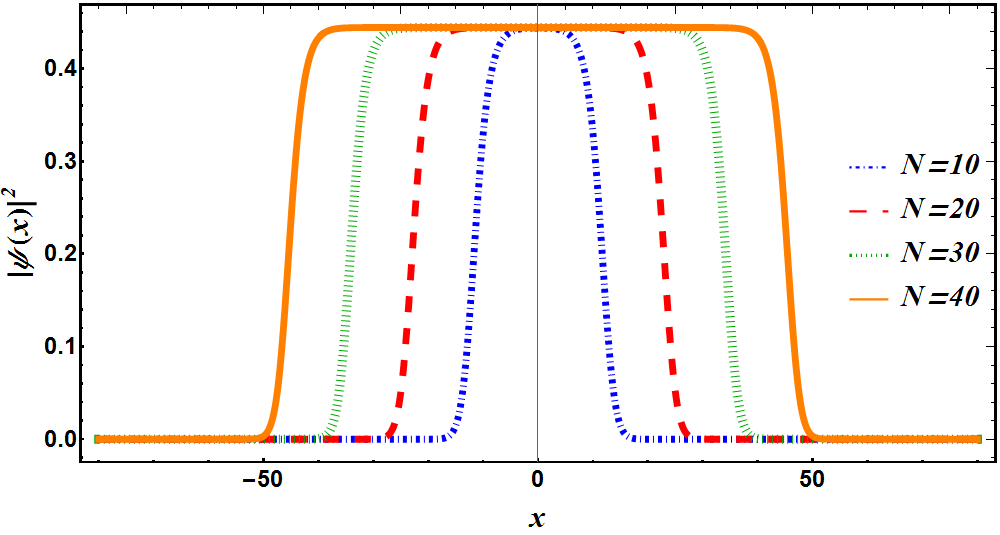} %
\caption{(Color Online) Spatial density distributions, $|\protect\psi (x)|^{2}$, as
given by the QD solution in Eq.(\ref{sol}), for different values of norm noted in Eq.(\ref{N}), as indicated in the figure.}
\label{profile}
\end{center}
\end{figure}
\begin{equation}
\phi (x)=-\frac{3\mu }{1+\sqrt{1+9\mu /2}\cosh \left( {\sqrt{-2\mu }x}%
\right) },  \label{sol}
\end{equation}%
where $\mu <0$ is the chemical potential, which takes values in a finite
bandgap,%
\begin{equation}
0<-\mu <2/9.  \label{2/9}
\end{equation}%

The norm of the QD state can be derived from Eq.(\ref{sol}) such that, \cite{Petrov2016}
\begin{equation}
N=\int_{-\infty }^{+\infty }\left\vert \psi (x)\right\vert ^{2}dx=\frac{4}{3}%
\left[ \ln \left( \frac{3\sqrt{-\mu /2}+1}{\sqrt{1+9\mu /2}}\right) -3\sqrt{%
-\mu /2}\right] .  \label{N}
\end{equation}%
Note that the dependence $N(\mu )$, as given by Eq. (\ref{N}), satisfies the
celebrated Vakhitov-Kolokolov (VK) criterion,
\begin{equation}
dN/d\mu <0,  \label{Vakh}
\end{equation}%
which is a necessary condition for the stability of localized states
supported by any self-attractive nonlinearity \cite{VK,Berge,Fibich}.

In the limit of $\mu \rightarrow -0$, the QD state (\ref{sol}) is similar to
traditional solitons dominated by the quadratic nonlinearity, such as ones
produced by the Korteweg -- de Vries equation \cite{KdV,Zakharov},%
\begin{equation}
\phi (x)\approx -\frac{3\mu }{2\cosh ^{2}\left( \sqrt{-\mu /2}x\right) }.
\label{KdV-like}
\end{equation}%
In the opposite limit, $\mu \rightarrow -2/9$ [see Eq. (\ref{2/9})], Eq. (%
\ref{sol}) produces quasi-flat droplets (\textquotedblleft puddles", in
terms of Ref. \cite{malomed}), with the nearly constant inner density,
\begin{equation}
\phi ^{2}(x)\leq n_{\mathrm{\max }}=4/9,  \label{nmax}
\end{equation}%
and a logarithmically large width,%
\begin{equation}
L\approx (3/2)\ln \left[ 1/\left( 1+9\mu /2\right) \right] .  \label{L}
\end{equation}%
Exactly at the border of the bandgap (\ref{2/9}), $\mu =-2/9$, QD (\ref{sol}%
) carries over into the front solutions,%
\begin{equation}
\phi _{\mathrm{front}}(x;\mu =-2/9)=\frac{2/3}{1+\exp \left( \pm 2x/3\right)
},  \label{front}
\end{equation}%
which connect the zero state and the constant-amplitude (alias
continuous-wave, CW) one,
\begin{equation}
\psi _{\mathrm{CW}}=(2/3)\exp ((2/9)it).  \label{CW}
\end{equation}%
In fact, the front solutions (\ref{front}) represent boundaries of a very
broad flat-top QD.

Density profiles of the QD solution (\ref{sol}) with different values of $N$
are plotted in Fig. \ref{profile}. In the figure, the profiles
corresponding to $N=20$, $30$, and $40$ clearly exhibit the flat-top shape.
In agreement with the prediction of the VK criterion, all the QDs (\ref{sol}%
) are stable solutions of Eq. (\ref{dbgp}).

The Galilean invariance of Eq. (\ref{dbgp}) [with $f(x)=0$] makes it
possible to set the QD in motion by application of the boost (kick) to them
with arbitrary parameter $k$, which determines the velocity of the moving
mode:%
\begin{equation}
\psi \left( x,t\right) =\exp \left[ ikx-i\left( k^{2}/2\right) t\right] \psi
\left( x-kt,t\right) .  \label{boost}
\end{equation}%
The effective mass of the moving QD coincides with its norm $N$, i.e., its
kinetic energy is%
\begin{equation}
E_{\mathrm{kin}}=N\left( k^{2}/2\right) .  \label{Ekin}
\end{equation}%
This result is used below to predict the threshold of the rebound of the
moving QD from a potential barrier, see Eq. (\ref{Uk}).

\section{Analytical results}\label{analytics}

\subsection{Localized modes pinned to narrow potential wells: exact solutions}
\begin{figure}
\begin{center}
\includegraphics[scale=0.50]{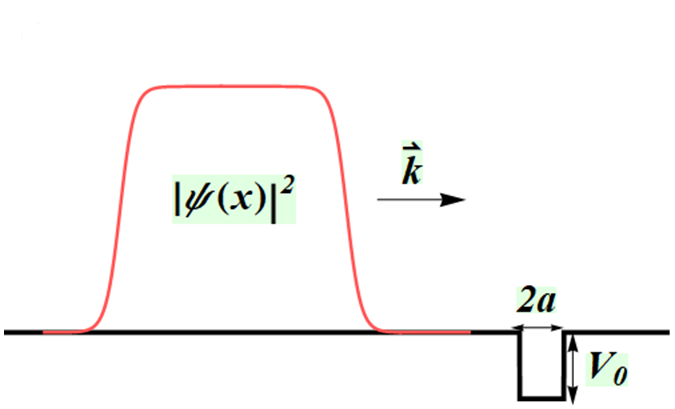}
\caption{(Color Online) A schematic
representation of a flat-top QD colliding with a rectangular potential well
with width $2a$ and depth $V_{0}$.}\label{cartoon}
\end{center}
\end{figure}

Our primary objective is to consider interactions of QDs with a potential
well/barrier, as described by Eq. (\ref{dbgp}) with $f(x)\neq 0$ and
illustrated by Fig. \ref{cartoon}. In this figure, the local potential is
taken as a rectangular well of width $2a$ and depth $V_{0}>0$:
\begin{equation}
f(x)=\left\{
\begin{array}{c}
-V_{0}\text{,}\,\,\,\,\text{for}\,-a<x<+a, \\
0\text{,}\,\,\,\,\text{for }|x|>a.%
\end{array}%
\right.  \label{barrier}
\end{equation}%
The same potential with $V_{0}<0$ represents a rectangular barrier of height
$-V_{0}$.

In this section we present analytical findings which for the case when the
finite well/barrier is approximated by a delta-function potential. The
approximation is valid provided that the width of the potential well is much
smaller than the size of the QD. In this limit, we obtain exact analytical
results which are later qualitatively corroborated by our numerical study.

Hence, we assume,
\begin{equation}
f(x)=f_{\varepsilon }(x)=-\varepsilon \delta (x),  \label{delta}
\end{equation}%
with strength
\begin{equation}
\varepsilon \equiv 2aV_{0}.  \label{epsilon}
\end{equation}%
The same potential (\ref{delta}) with $\varepsilon <0$, i.e., $V_{0}<0$ in
Fig. \ref{cartoon}, represents a narrow potential barrier, such as the one
used as the soliton splitter in interferometers \cite%
{Martin,Ahufinger,Helm2,Cornish2,Gardiner}.
\begin{figure}[tbp]
\begin{center}
\includegraphics[scale=0.25]{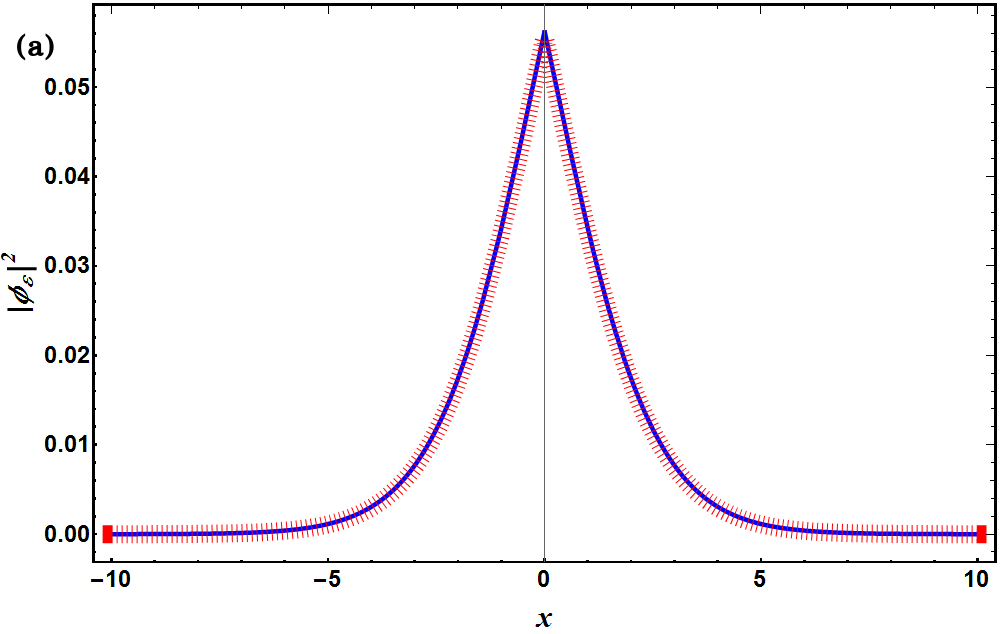}\\
\includegraphics[scale=0.25]{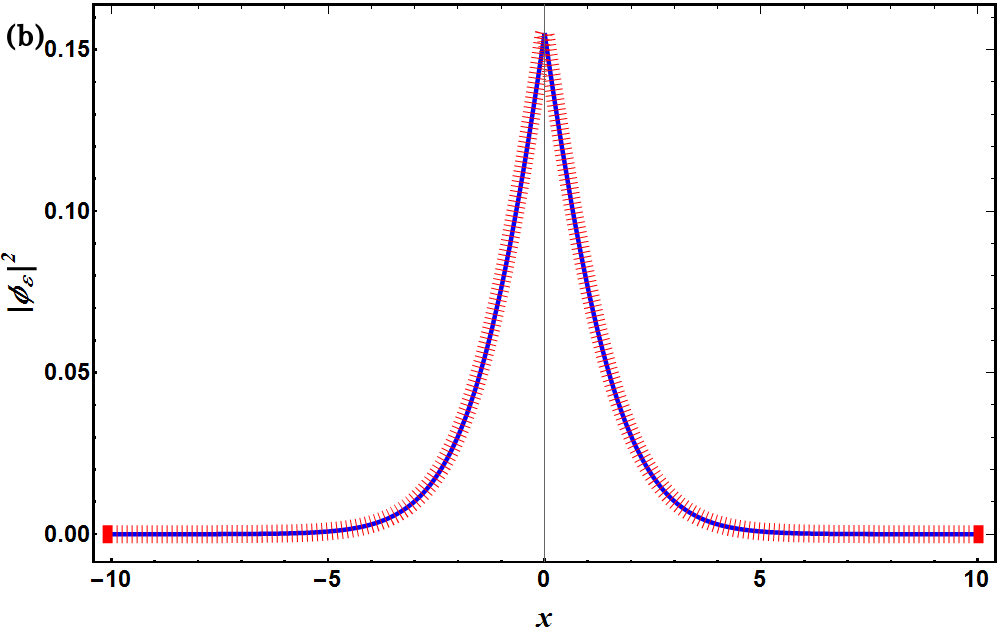}\\
\includegraphics[scale=0.25]{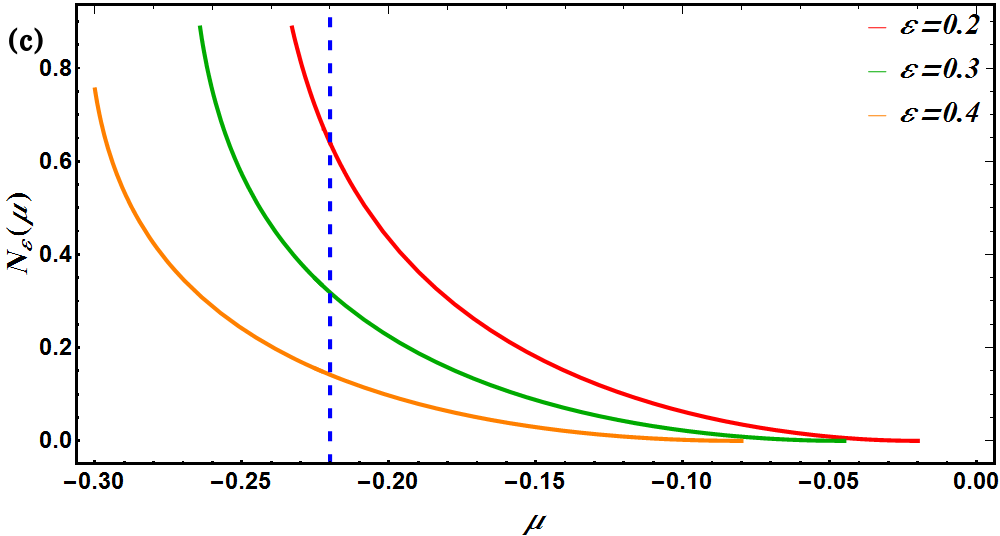}
\caption{(Color Online) Comparison between typical examples of analytical solutions (blue
solid lines) given, respectively, by Eqs. (\protect\ref{exact}), (\protect
\ref{xi}) and (\protect\ref{exact2}), (\protect\ref{eta}) with their
counterparts (red-bar lines), produced by the numerical solution of Eq. (%
\protect\ref{stationary}) [with the delta-function replaced by its Gaussian
approximation (\protect\ref{Gauss})], is depicted in panels (a) and (b). The
respective values of $\protect\varepsilon $ in (a) and (b) are $0.2$ and $0.3
$. It is seen that the analytical and numerical solutions are
indistinguishable. (c) The continuous dependence $N_{\protect\varepsilon }(%
\protect\mu )$, as produced by the analytical solutions given, severally, by
Eqs. (\protect\ref{exact}) and (\protect\ref{xi}) at $\protect\mu >-2/9$,
and by Eqs. (\protect\ref{exact2}) and (\protect\ref{eta}) at $\protect\mu %
<-2/9$, for $\protect\varepsilon =0.2$, $0.3$ and $0.4$ (red, green, and
orange lines, respectively). The vertical blue dashed line indicates the
boundary between the two analytical solutions, $\protect\mu =-2/9$. At $%
\protect\mu <-2/9$, the curves are extended very close to the right edge of
interval (\protect\ref{annex}), $-\protect\mu =2/9+\protect\varepsilon ^{2}/2
$, beyond which solution (\protect\ref{exact2}) does not exist. Counterparts
of these curves produced by the numerical solution (not shown here) are
virtually identical to the analytical ones.}
\label{vkprofile}
\end{center}
\end{figure}

Exact analytical solutions for nonlinear modes pinned to the delta-function
potential well defined by Eq. (\ref{delta}) are produced below. We
substantiate these analytical results by comparison with their numerically
found counterparts, as shown in Fig. \ref{vkprofile}(a,b). The respective
numerical solutions of Eq. (\ref{stationary}) were produced with the $\delta
$-potential approximated by means of the standard Gaussian,
\begin{equation}
\tilde{\delta}(x)=\frac{1}{\sqrt{\pi }\sigma }\exp \left( -\frac{x^{2}}{%
\sigma ^{2}}\right) ,  \label{Gauss}
\end{equation}%
with a small width, $\sigma =0.03$ \cite{appdelta}.

Exact solutions of Eq. (\ref{stationary}) for QDs pinned to the
delta-function potential well defined by Eq. (\ref{delta}), one can
construct it as a direct extension of its free-space counterpart, given by
Eq. (\ref{sol}), \textit{viz}.,
\begin{equation}
\phi _{\varepsilon }(x,\mu )=-\frac{3\mu }{1+\sqrt{1+9\mu /2}\cosh \left[
\sqrt{-2\mu }\left( {|x|+\xi }\right) \right] },  \label{exact}
\end{equation}%
in the interval (\ref{2/9}) of the values of $\mu $, where the positive
\textquotedblleft lost length" (i.e., the shift of the maximum from its position in the absence of the attractive potential) is%
\begin{equation}
2\xi =\sqrt{-\frac{2}{\mu }}\ln \left[ \frac{\sqrt{\varepsilon ^{2}+\left(
1+9\mu /2\right) \left( -2\mu -\varepsilon ^{2}\right) }+\varepsilon }{\sqrt{%
1+9\mu /2}\left( \sqrt{-2\mu }-\varepsilon \right) }\right]  \label{xi}
\end{equation}%
(this way of constructing solutions for QDs pinned to the delta-function
potential was proposed in Ref. \cite{abdullaev}). It is seen from Eq. (\ref%
{xi}) that the exact solutions for the pinned states exists in the following
interval of values of the chemical potential,%
\begin{equation}
\varepsilon ^{2}/2<-\mu <2/9,  \label{reduced}
\end{equation}%
which is reduced in comparison to the bandgap mentioned in Eq. (\ref{2/9}).
Thus it follows from Eq. (\ref{reduced}) that the pinned states of the QDs
exist only if the delta-function pinning potential is not too strong,
\textit{viz}.,
\begin{equation}
\varepsilon <\varepsilon _{\max }=2/3.  \label{2/3}
\end{equation}%
A majority of numerical results are reported below for $\varepsilon \equiv
V_{0}<2/3$, i.e., condition (\ref{2/3}) holds for them.

In the limit case of $\mu =-2/9$, the exact solution for the pinned QD,
given by Eqs. (\ref{exact}) and (\ref{xi}) takes a simple but nontrivial
form:%
\begin{equation}
\phi _{\varepsilon }(x,\mu =-2/9)=\frac{2/3}{1+\varepsilon \left(
2/3-\varepsilon \right) ^{-1}\exp \left( 2|x|/3\right) }.  \label{limit}
\end{equation}%
Note that, while the exact solution in Eq. (\ref{sol}) with $\mu =-2/9$
degenerates into the CW state (\ref{CW}) with the divergent norm, solution (%
\ref{limit}) with the same $\mu $ remains a localized one, with a final norm
[defined as per Eq. (\ref{N})]%
\begin{equation}
N_{\varepsilon }\left( \mu =-2/9\right) =\frac{4}{3}\ln \left( \frac{2}{%
3\varepsilon }\right) -2\left( \frac{2}{3}-\varepsilon \right) .
\label{N(2/9)}
\end{equation}

In the opposite limit corresponding to the left edge of interval (\ref%
{reduced}),
\begin{equation}
\mu \rightarrow \mu _{0}\equiv -\varepsilon ^{2}/2,  \label{mu0}
\end{equation}%
the solution given by Eqs. (\ref{exact}) and (\ref{xi}) degenerates into
\begin{equation}
\phi _{\varepsilon }(x;\mu =-\varepsilon ^{2}/2)=\mathcal{A}\exp \left(
-\varepsilon |x|\right) ,  \label{a}
\end{equation}%
where $\mathcal{A}$ is an arbitrary infinitesimal amplitude. This limit form
of the solution is tantamount to the commonly known bound state maintained
by the delta-function potential well in the framework of the linear Schr\"{o}%
dinger equation.

A typical example of the exact solution produced by Eqs. (\ref{exact}) and (%
\ref{xi}) is displayed in Fig. \ref{vkprofile}(a), where it is compared to
its counterpart produced by a numerical solution of Eq. (\ref{stationary})
with the potential well taken as per Eqs. (\ref{barrier}) and (\ref{epsilon}%
). It is seen that the analytical and numerical solutions are virtually
identical ones.

The comparison of the value of the norm given by Eq. (\ref{N(2/9)}) and the
infinitesimal norm at $\mu =-\varepsilon ^{2}/2$ corresponding to Eq. (\ref%
{a}) suggests that the dependence $N_{\varepsilon }(\mu )$ is monotonously
decreasing. This feature, which is corroborated by Fig. \ref{vkprofile}(c),
suggests that the family of pinned QDs are stable, according to the
above-mentioned VK criterion (\ref{Vakh}). \cite{VK,Berge,Fibich}. Their
stability is indeed corroborated by direct simulations of Eq. (\ref{dbgp})
(not shown here in detail, as results of the simulations are
straightforward).

In the presence of the attractive delta-functional potential, it is possible
to find another exact solution for pinned QDs in the region of $\mu <-2/9$,
where free-space solitons (\ref{sol}) \emph{do not exist}. In the case when $%
\varepsilon $ satisfies condition (\ref{2/3}), this region extends the
existence interval (\ref{2/9}). The corresponding exact solution, which has
no counterparts in the free space, is%
\begin{equation}
\tilde{\phi}_{\varepsilon }(x)=-\frac{3\mu }{1+\sqrt{-(1+9\mu /2)}\sinh %
\left[ \sqrt{-2\mu }\left( {|x|+\eta }\right) \right] }  \label{exact2}
\end{equation}%
[cf. expression (\ref{sol}) for the solution constructed above], the
positive \textquotedblleft lost length" is%
\begin{equation}
2\eta =\sqrt{-\frac{2}{\mu }}\ln \left[ \frac{\sqrt{\varepsilon ^{2}-\left(
1+9\mu /2\right) \left( \varepsilon ^{2}+2\mu \right) }+\varepsilon }{\sqrt{%
-\left( 1+9\mu /2\right) }\left( \sqrt{-2\mu }-\varepsilon \right) }\right] ,
\label{eta}
\end{equation}%
cf. Eq. (\ref{xi}). Expression (\ref{exact2}) is relevant if it yields real
values of $\eta $. This condition holds in the following finite interval of
values of the chemical potential,%
\begin{equation}
2/9\leq -\mu \leq 2/9+\varepsilon ^{2}/2.  \label{annex}
\end{equation}%
It is a straightforward annex to interval (\ref{reduced}) in which the above
solution (\ref{exact}) exists. As well as the latter solution, the one given
by Eqs. (\ref{exact2}) and (\ref{eta}) satisfies the VK stability criterion (%
\ref{Vakh}), as shown in Fig. \ref{vkprofile}(c). Note that, in the limit of
$\mu +2/9\rightarrow -0$, solution (\ref{exact2}) takes the same form (\ref%
{limit}) to which the above solution (\ref{exact}) amounts at $\mu
+2/9\rightarrow +0$, hence solution (\ref{exact2}) provides smooth
continuation of its counterpart (\ref{exact}) across the point of $\mu
+2/9=0 $ [see an illustration in Fig. \ref{vkprofile}(c)].

The exact solution (\ref{exact2}) can also be constructed in the region of%
\begin{equation}
\varepsilon >2/3,  \label{>2/3}
\end{equation}%
i.e., outside of the region (\ref{2/3}). In this case, expression (\ref{eta}%
) is replaced by%
\begin{equation}
\eta =\frac{1}{\sqrt{-2\mu }}\ln \left[ \frac{\varepsilon -\sqrt{\varepsilon
^{2}-\left( 1+9\mu /2\right) \left( \varepsilon ^{2}+2\mu \right) }}{\sqrt{%
-\left( 1+9\mu /2\right) }\left( \sqrt{-2\mu }-\varepsilon \right) }\right] ,
\label{eta2}
\end{equation}%
and the solution exists in the same interval of the chemical potential as
given by Eq. (\ref{annex}), while it has no counterpart at $-\mu <2/9$.
Indeed, in the limit of $\mu +2/9\rightarrow -0$ Eqs. (\ref{exact2}) and (%
\ref{eta2}) demonstrate that the solution degenerates into $\phi =0$ [on the
contrary to the nonzero solution (\ref{limit}), into which solution (\ref%
{exact}) carries over in the limit of $\mu +2/9\rightarrow +0$], thus
providing the natural continuity with the absence of the pinned solution at $%
\mu >-2/9$ in the region (\ref{>2/3}). On the other hand, point (\ref{mu0}),
where solution (\ref{exact}) vanishes, as shown above [see Eq. (\ref{a})],
is now an internal point of interval (\ref{annex}), due to condition (\ref%
{>2/3}). At this point, the present solution takes a relatively simple form,%
\begin{eqnarray}  \label{non-vanishing}
&&\tilde{\phi} _{\varepsilon }(x;\mu =-\varepsilon ^{2}/2)  \notag \\
&&=\frac{3\varepsilon ^{2}/2}{1+\sqrt{9\varepsilon ^{2}/4-1}\sinh \left[
\varepsilon |x|+(1/2)\ln \left( 9\varepsilon ^{2}/4-1\right) \right] }.
\notag \\
\end{eqnarray}

The family of the exact solutions produced by Eqs. (\ref{exact2}) and (\ref%
{eta2}) also satisfies the VK criterion; in particular, the norm decreases
from a finite value, corresponding to solution (\ref{non-vanishing}), at $%
\mu =-\varepsilon ^{2}/2$ to $N=0$ at $\mu =-2/9$. The compliance of the
present family with the VK criterion implies its stability.

\subsection{The Thomas-Fermi (TF) approximation for pinned state}

A specific analytical approximation for solutions to Eq. (\ref{stationary})
with the deep potential well (\ref{barrier}), i.e., with large $V_{0}$ and $%
\mu <0$, is provided by the TF approximation, which neglects the derivative
(kinetic-energy) term:%
\begin{equation}
\phi _{\mathrm{TF}}(x)=\left\{
\begin{array}{c}
\frac{1}{2}+\sqrt{\frac{1}{4}-\left( V_{0}-\mu \right) },~\mathrm{at~}|x|<a,
\\
\frac{1}{2}+\sqrt{\frac{1}{4}+\mu },~\mathrm{at~}|x|>a%
\end{array}%
\right.   \label{TF}
\end{equation}%
[in fact, the solution (\ref{TF}) at $|x|>a$ is relevant for the flat-top
QD, with $\mu \approx -2/9$, the nearly constant solution around the
potential well being close to the CW background (\ref{CW})]. Then, for a
small perturbation $\delta \phi $ added to $\phi _{\mathrm{TF}}(x)$, the
linearization of the nonlinear term in Eq. (\ref{stationary}) produces, in
the lowest approximation with respect to the fact that $V_{0}$ is a large
parameter, an effective term in the GPE, $+3\varepsilon \delta (x)\cdot
\delta \phi $, where $\varepsilon $ is defined as per Eq. (\ref{epsilon}),
and $\delta (x)$ is introduced in the same approximation as in Eq. (\ref%
{delta}). Then, the combination of this term and one corresponding to Eq. (%
\ref{delta}) gives rise to an effective \emph{repulsive} potential,
\begin{equation}
f_{\mathrm{effective}}(x)=+2\varepsilon \delta (x).  \label{effective}
\end{equation}%
This crude analysis predicts, in a quantitative form, that the collision of
a flat-top QD with the deep potential well may lead, instead of the
naturally expected passage, to the counter-intuitive outcome in the form of
\emph{rebound}. This possibility is indeed demonstrated by numerical results
reported below. It is relevant to mention that reflection of incident wave
packets from a potential well is a well-known quantum effect, which was also
elaborated in the framework of the MF theory \cite{lee,brand}.

\subsection{The interaction of the moving QD with a narrow potential barrier}
As mentioned above, collisions of moving matter-wave packets with narrow
potential barriers is a problem of high relevance for the design of soliton
interferometers \cite{Martin,Ahufinger,Helm2,Cornish2,Gardiner}. In the
present context, treating the repulsive delta function potential (\ref{delta}%
) as a perturbation \cite{old} makes it possible to predict the threshold
value $k_{\mathrm{thr}}$ of velocity $k$ [see Eq. (\ref{boost})] which
separates the rebound and passage of the incoming QD. In the adiabatic
approximation, which neglects deformation of the QD under the action of the
barrier, an effective potential of the interaction of the QD with the
barrier can be found, in the adiabatic approximation (which neglects
deformation of the QD due to its interaction with the local potential) as%
\begin{eqnarray}  \label{Xi}
&&U_{\mathrm{eff}}(\Xi )=-\varepsilon \int_{-\infty }^{+\infty }\delta
(x)\left\vert \psi \left( x-\Xi \right) \right\vert ^{2}dx  \notag \\
&&=-\frac{9\varepsilon \mu ^{2}}{\left[ 1+\sqrt{1+9\mu /2}\cosh \left( \sqrt{%
-2\mu }\Xi \right) \right] ^{2}},  \notag \\
\end{eqnarray}%
where $\Xi (t)$ is the coordinate of the center of the moving QD. Obviously,
the height of the effective potential barrier (\ref{Xi}) is%
\begin{equation}
U_{\max }=\frac{9\varepsilon \mu ^{2}}{\left( 1+\sqrt{1+9\mu /2}\right) ^{2}}%
,  \label{Umax}
\end{equation}%
hence the comparison of this expression with the QD's kinetic energy (\ref%
{Ekin}) predicts the threshold value as%
\begin{equation}
k_{\mathrm{thr}}^{2}=\frac{18\varepsilon \mu ^{2}}{N(\mu )\left( 1+\sqrt{%
1+9\mu /2}\right) ^{2}},  \label{thr}
\end{equation}%
where $N(\mu )$ is to be taken as per Eq. (\ref{N}). The incident QD with $%
k<k_{\mathrm{thr}}$ or $k>k_{\mathrm{thr}}$ is expected, respectively, to
bounce back or pass over the barrier.

In the limit case of flat-top QDs, which correspond to $\mu $ close to $-2/9$
[see Eq. (\ref{2/9})], Eqs. (\ref{Umax}) and (\ref{thr}) take a simplified
form:%
\begin{equation}
U_{\max }\approx \frac{4}{9}\varepsilon ,~k_{\mathrm{thr}}^{2}\approx \frac{%
8\varepsilon }{9N}.  \label{Uk}
\end{equation}%
It may also be relevant to consider shuttle motion of the QD in a cavity
formed by two narrow barriers, separated by a large distance $L$. The period
of the motion with velocity $k$ is $T=2L/k$, hence the largest shuttle
frequency which can be maintained by the cavity is $\omega _{\max }=\pi k_{%
\mathrm{thr}}/L$.
\begin{figure}[tbp]
\begin{center}
\includegraphics[scale=0.2]{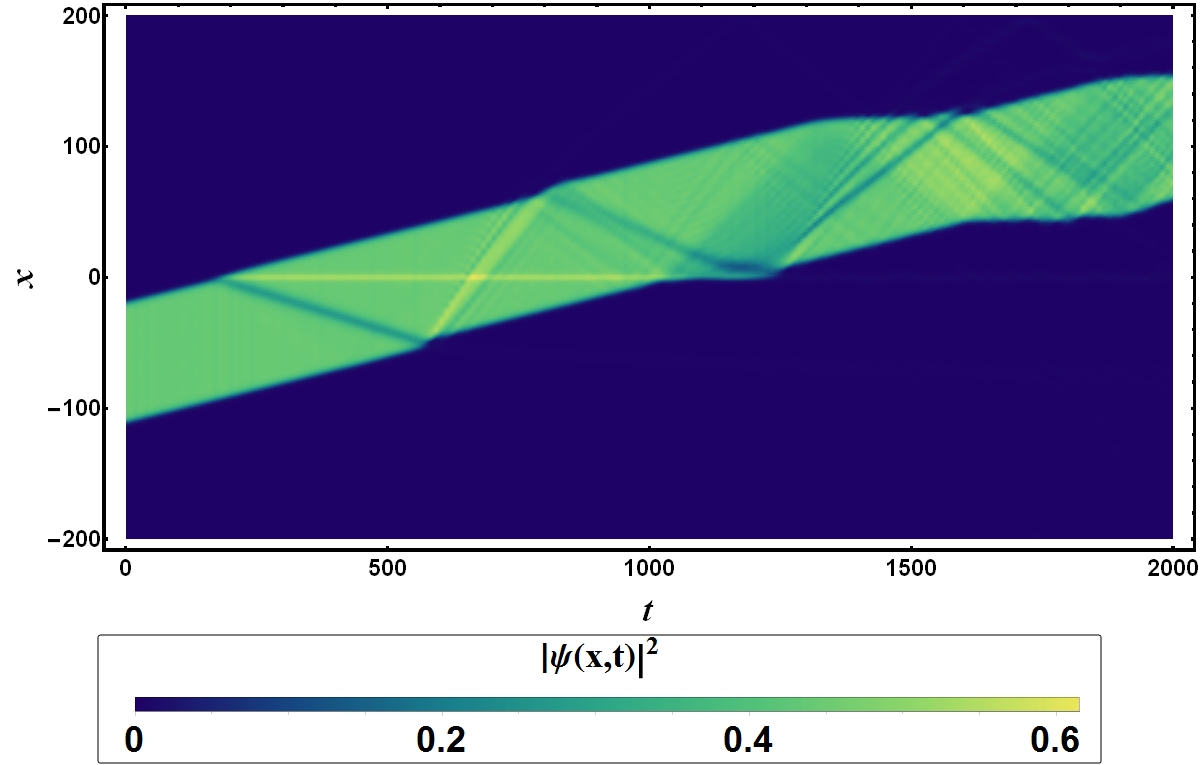}
\caption{(Color Online) The plot of the spatio-temporal evolution of the density, $%
\left\vert \protect\psi \left( x,t\right) \right\vert ^{2}$, presents an
example of the full passage of the incident QD through the potential well, for $V_0=0.13$, $N=40$ and $k=0.1$.}
\label{total_trans}
\end{center}
\end{figure}

\section{Numerical analysis}\label{numerics}
 Following the analytical considerations presented in the
previous section and compared to the numerical results in Fig. \ref%
{vkprofile}, in this section we aim to study the dynamics of QDs interacting
with the potential well/barrier. To this end, simulations of Eq. (\ref{dbgp}%
) were performed by means of the split-step method based on the fast Fourier
transform.
\begin{figure}[h!]
\includegraphics[scale=0.2]{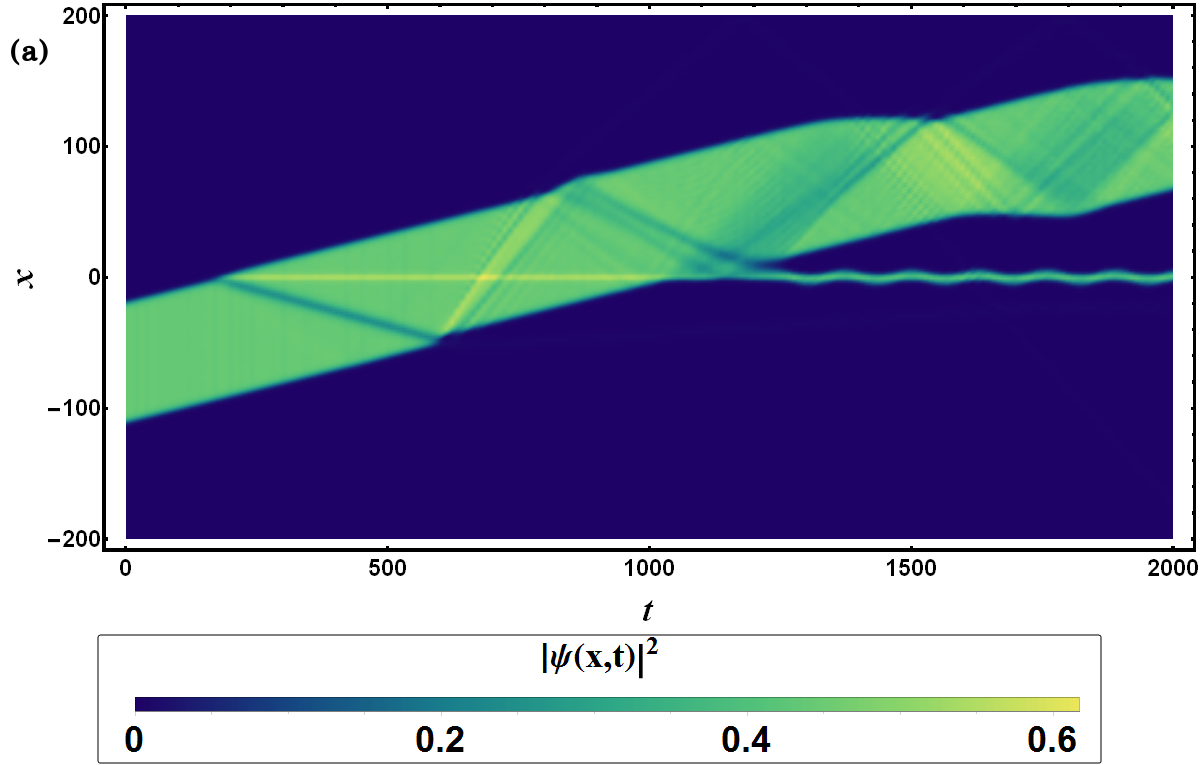}\\
\vfill\vfill\vfill\vfill
\includegraphics[scale=0.25]{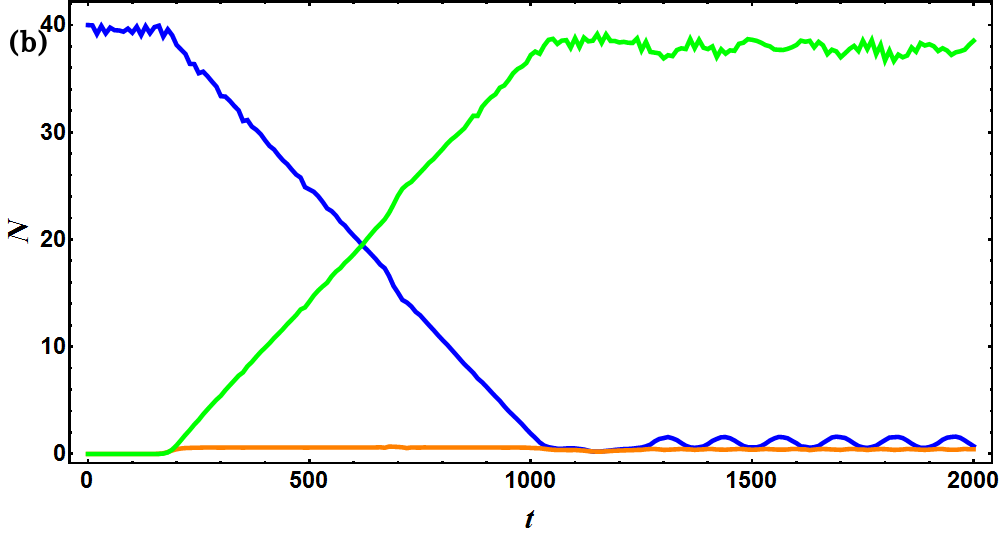}
\caption{(Color Online) (a) An example of the partial transmission of the incident QD
through the potential well [as noted in Eq.(\protect\ref{barrier})] with
depth $V_{0}=0.14$. A very small share of the wave function is trapped in
the well. (b) The temporal evolution of the norm for the transmitted (green
solid line), trapped (orange solid line) and reflected (blue solid line)
shares, defined as per Eqs. (\protect\ref{NNN}).}
\label{partial_trans}
\end{figure}

The numerical results are reported for the rectangular potential (\ref%
{barrier}) with width $2a=1$, which is small in comparison with other length
scales, such as the width of the incident QDs (cf. Fig. \ref{profile}), and
justifies the comparison with approximation (\ref{delta}). In this case, Eq.
(\ref{epsilon}) yields $\varepsilon =V_{0}$. We use zero boundary condition
corresponding to an infinitely deep potential box occupying a broad spatial
domain, $|x|<L=200$. The initial wave function was taken as
\begin{equation}
\psi (x,t=0)=\exp (ikx)\phi (x-x_{\mathrm{init}}),  \label{x_init}
\end{equation}%
where $\phi (x)$ is the stationary solution (\ref{sol}) for the QD, $k$ is
the kick which sets the QD in motion, and $x_{\mathrm{init}}$ is its initial
position.

\subsection{Moving droplets}

It is natural to expect that the travelling droplets colliding with the
narrow potential well or barrier may demonstrate transmission or reflection,
or even trapping, depending on the initial kick ($k$) in Eq. (\ref{x_init})
and the strength of the potential. Splitting of the incident QD in
transmitted, reflected, and trapped fragments may be expected too.

Norms of the transmitted, trapped, and reflected components of the wave
function are defined as,
\begin{eqnarray}
N_{\mathrm{trans}}&=&\int_{a}^{L}{dx|\psi (x,t)|^{2}},\nonumber\\N_{\mathrm{trap}%
}&=&\int_{-a}^{+a}{dx|\psi (x,t)|^{2}},\nonumber\\N_{\mathrm{ref}}&=&\int_{-L}^{-a}{dx|\psi
(x,t)|^{2},}  \label{NNN}
\end{eqnarray}%
which are subjected to the obvious constraint, $N_{\mathrm{trans}}+N_{\mathrm{%
trap}}+N_{\mathrm{ref}}=N$. 

First, we check effects of variation of the potential's strength for fixed $k
$ and $N$. The simulations started with the QD placed at $x_{\mathrm{init}%
}=-65$ [see Eq. (\ref{x_init})] and moving with velocity $k>0$ towards the
local potential barrier/well defined by Eq. (\ref{barrier}), with the width
scaled to be $2a=1$, as said above.

\begin{figure}[tbp]
\includegraphics[scale=0.2]{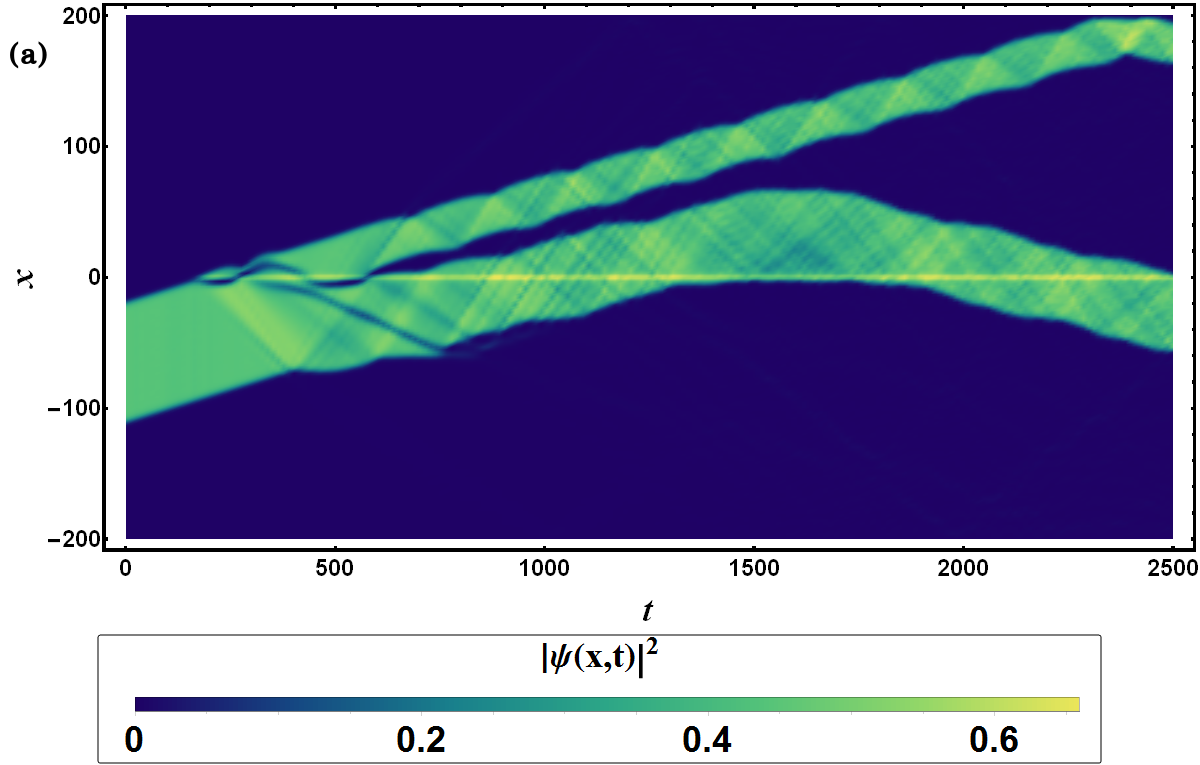} \\
\vfill\vfill\vfill\vfill
\includegraphics[scale=0.25]{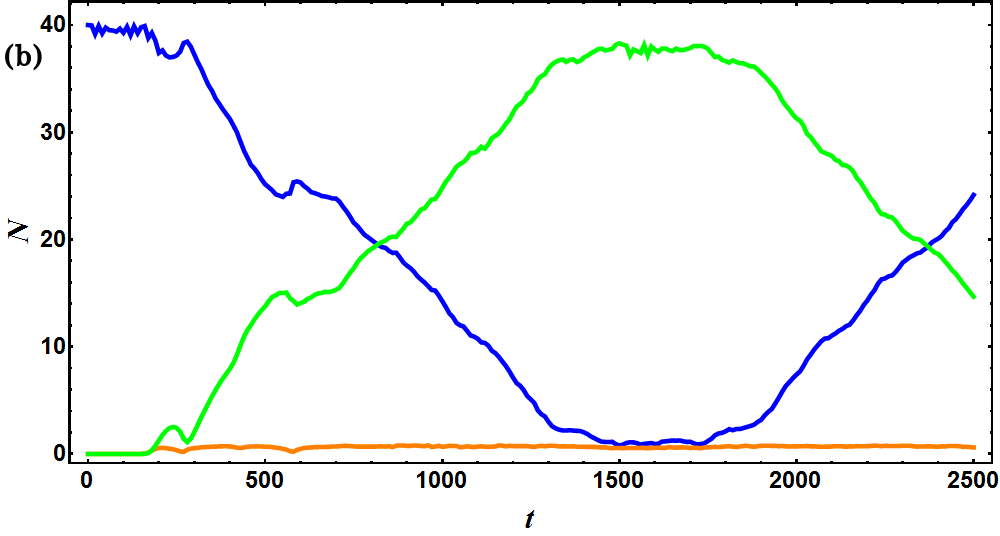}
\caption{(Color Online) (a) An example of the partial transmission and partial reflection
of the incident QD colliding with the potential well, whose depth is $V_{0}=0.248$. (b) The temporal evolution of the
transmitted (green solid line), trapped (orange solid line), and reflected (blue solid line) norms as defined in Eq. (\protect\ref%
{NNN}).}
\label{partial_ref}
\end{figure}

\begin{figure}[tbp]
\includegraphics[scale=0.2]{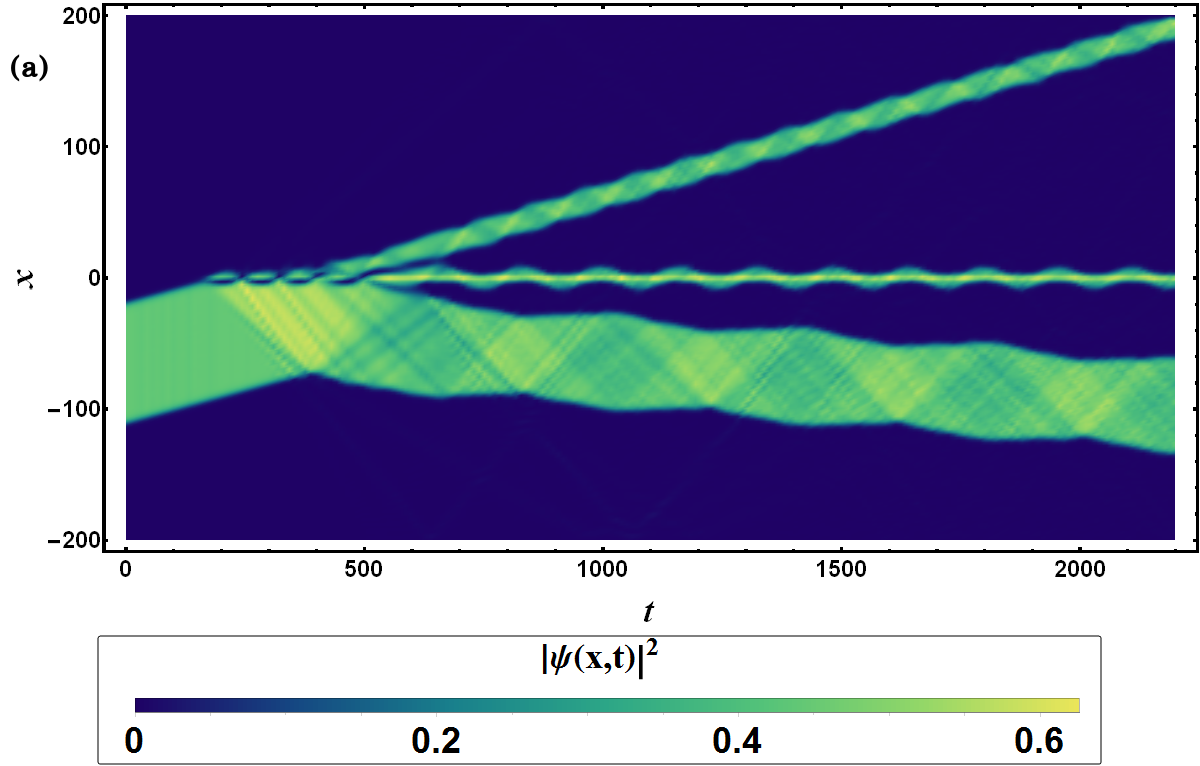} \\
\vfill\vfill\vfill\vfill
\includegraphics[scale=0.25]{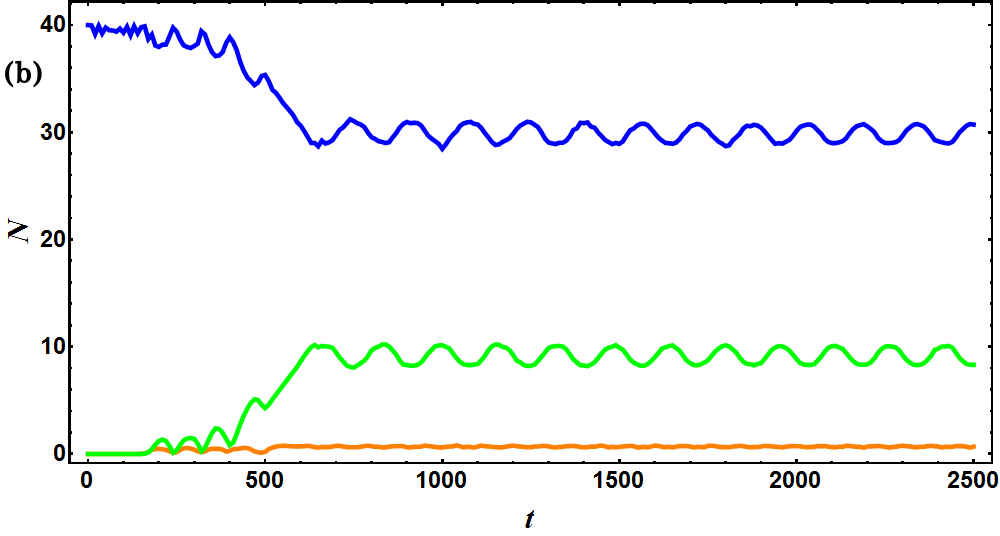} \\
\vfill\vfill\vfill\vfill
\includegraphics[scale=0.25]{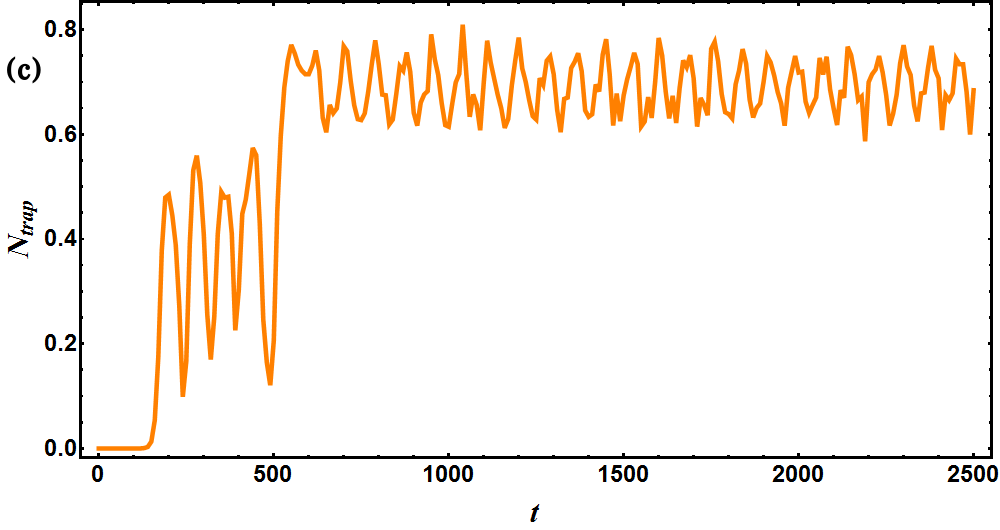}
\caption{(Color Online) (a) The splitting of the QD impinging onto potential well, with depth $V_{0}=0.3$, into the transmitted, trapped and
reflected parts are depicted. (b) Description of the time evolution of the norm for transmitted (green solid line), trapped (orange solid line) and reflected waves (blue solid line). 
(c) The figure focuses on the evolution of the
norm corresponding to the trapped wave. The complementary energy scenario is captured in Fig.~\ref{total_energy}}
\label{3_part}
\end{figure}

\begin{figure}[tbp]
\begin{center}
\includegraphics[scale=0.3]{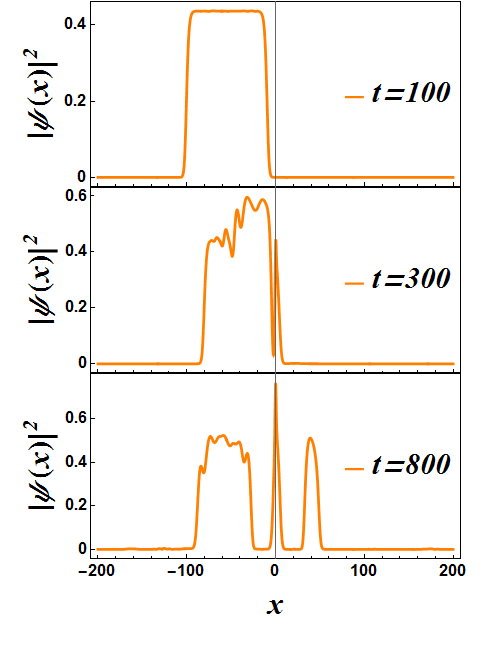}
\caption{(Color Online) Snapshots of the QD colliding with the potential well of depth $
V_{0}=0.3$ at $t=100$ (top panel), $t=300$ (middle panel), $t=800$ (bottom panel). Over the time the incident QD splits
into a larger reflected part and a smaller transmitted one, along with a
narrow trapped mode as described in the bottom panel.}
\label{snap_3_part}
\end{center}
\end{figure}

\begin{figure}[tbp]
\includegraphics[scale=0.2]{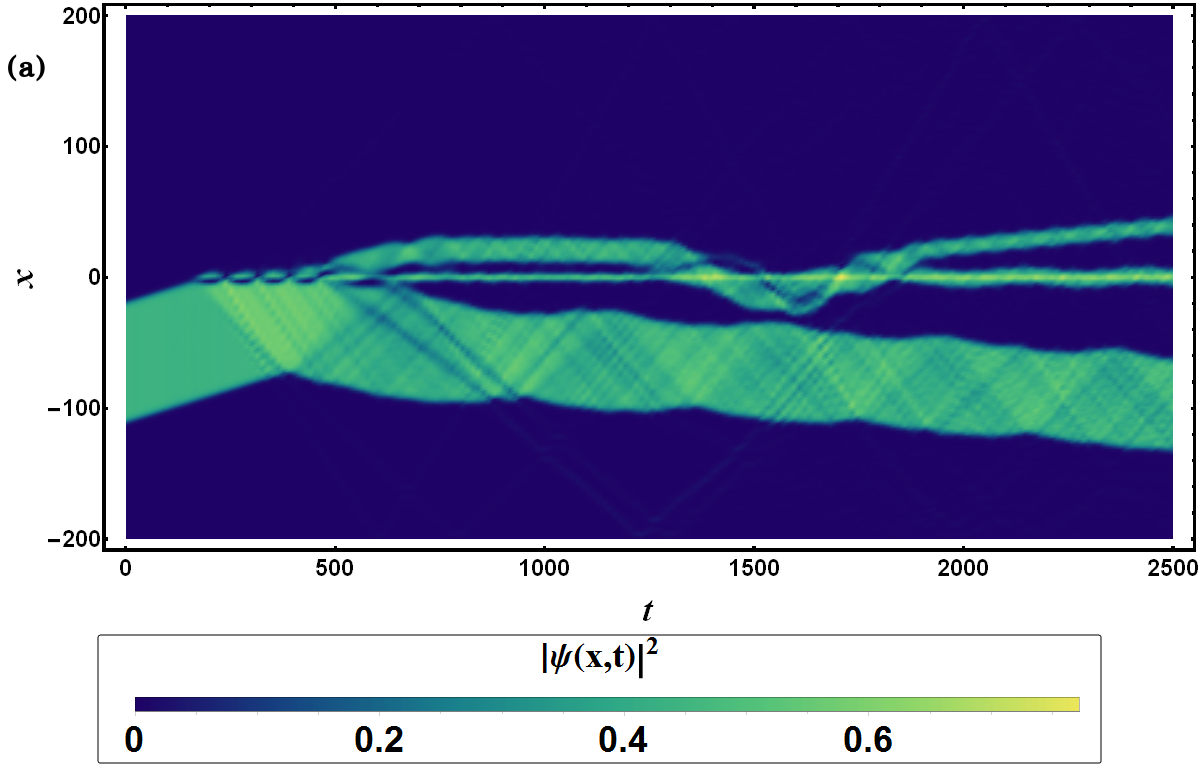}\\ %
\vfill\vfill\vfill\vfill
\includegraphics[scale=0.25]{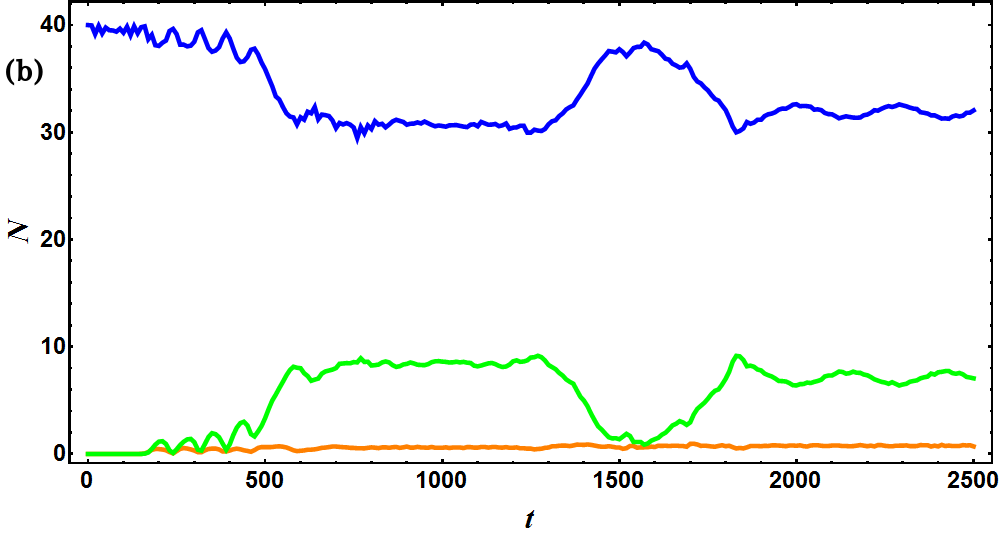}
\caption{(Color Online) (a) The partial reflection at $V_{0}=0.31$, with a very small
packet trapped in the well. (b) The variation of the transmitted (green solid line), trapped (orange solid line)
and reflected (blue solid line) norms are noted which clearly indicates a predominantly reflection of the waves.}
\label{half_ref}
\end{figure}

First, we present characteristic results obtained for the total norm $N=40$
and velocity $k=0.1$. We keep these values, unless mentioned otherwise. In
particular, this relatively low collision velocity is appropriate for
producing nontrivial results, while the fast moving QD simply passes the
barrier/well, or elastically bounces back from a very tall barrier.
Conclusions valid for generic values of the $N$ and $k$ are presented below.

Figure \ref{total_trans} displays an example of the total transmission of
the QD through the potential well (\ref{barrier}) with depth $V_{0}=0.13$.
Yellow lines in the density plot indicate excitations inside of the QD
generated by the collision with the well (intrinsic excitation modes in
quasi-1D QDs were studied in Ref. \cite{Tylutki}).For values of the
parameters corresponding to this figure, Eq. (\ref{Uk}) yields $k_{\mathrm{%
thr}}\approx 0.05$. The fact that this value is only half of the one used
here, $k=0.1$, immediately explains the total transmission observed in this
case (the same argument explains the transmission in Fig. \ref{partial_trans}%
).

A trapped fraction of the wave field appears for a larger depth of the
potential well. As an example, Fig. \ref{partial_trans} shows the
transmission with partial capture of the wave function at $V_{0}=0.14$. The
small-amplitude narrow trapped component observed in this case is
approximately described by solution (\ref{a}) of the linearized equation.
Further increase of $V_{0}$ causes partial reflection of the QD from the
potential well, which is in qualitative agreement with the above prediction,
see Eq. (\ref{effective}). The trapped share remains very small, as seen in
Fig. \ref{partial_ref} at $V_{0}=0.248$. The partial reflection may be
interpreted in a two-mode approximation \cite{ernst}, with the modes representing a small
fraction of the droplet trapped in the well and the large moving one. When
the two modes overlap, this may result in mutual repulsion when they are out
of phase. Thus, when the QD is situated close to the narrow-potential's
center, it may be reflected if the repulsion from the trapped mode overcomes
the attraction force directly exerted by the potential.

\begin{figure}[tbp]
\includegraphics[scale=0.2]{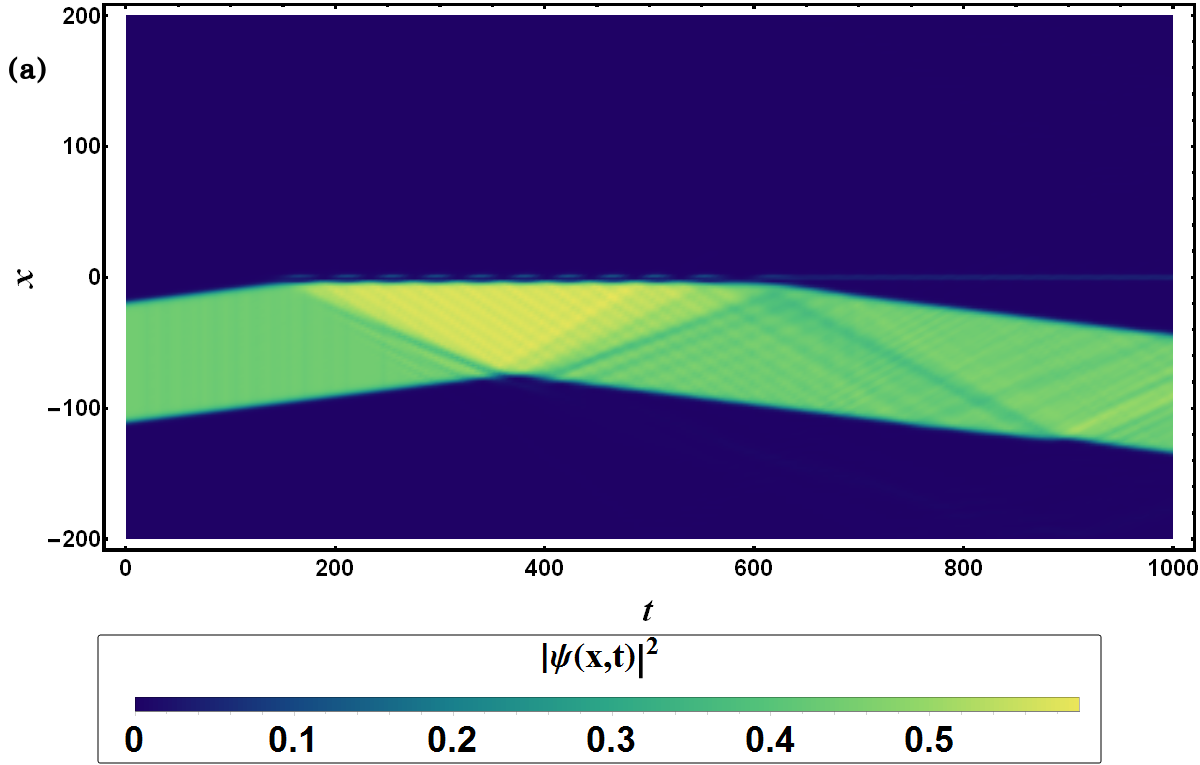} \\%
\vfill\vfill\vfill\vfill
\centering{\includegraphics[scale=0.33]{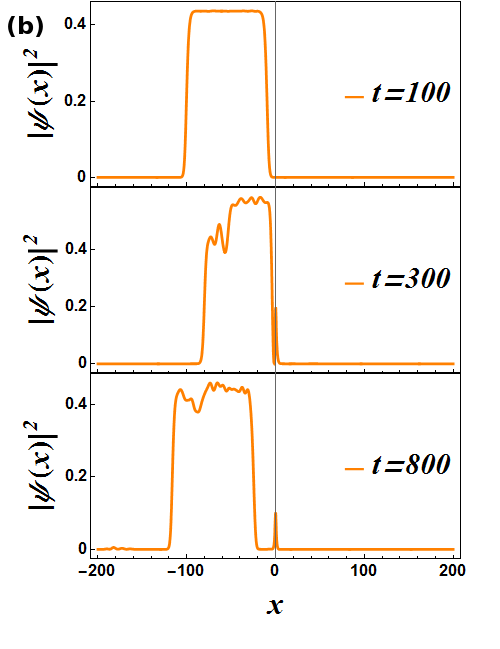}}
\caption{(Color Online) (a) A reasonably deep potential well with depth $V_{0}=0.55$ leads to nearly total reflection of the QD.
(b) Snapshots of the density profile at
different times [$t=100$ (top panel), $t=500$ (middle panel) and $t=800$ (bottom panel)] clearly substantiates the observation of (a). However, a very small fraction of trapped 
states can also be viewed from (a) and in the bottom panel of (b).}
\label{total_ref}
\end{figure}

The further increase of the depth of the potential well leads to a regime in
which the incident QD splits in three different parts, \textit{viz}., the
transmitted, reflected, and (small) trapped wave packets. Figures \ref%
{3_part} and \ref{half_ref} display examples of this outcome for depths $%
V_{0}=0.30$ and $V_{0}=0.31$, respectively. The former case is additionally
illustrated in Fig. \ref{snap_3_part} by a set of snapshots of the density
profiles. It is seen that all the three packets are produced by the
collision is in excited states, and a small change of $V_{0},$ from $0.30$
to $0.31$, leads to a conspicuous change of the dynamical picture. In terms
of energy, these collisions are considered in the next subsection.

In accordance with the above qualitative analysis, based on the TF
approximation and Eq. (\ref{effective}), nearly full reflection of the
incident QD is observed at larger values of the well's depth $V_{0}$, as
illustrated in Figs. \ref{total_ref} for $V_{0}=0.55$. A very small trapped
component is observed too, while transmission ceases to exist.

It is relevant to note that, besides the collision, splitting of the 1D QD
into two or several fragments can also be caused by imprinting sufficiently
small intrinsic excitation onto it \cite{malomed}. Thus, flat-top QDs, even
if they are stable solutions against small perturbations, are subject to
internal fragility.

\subsection{Analysis of energy in the trapping process}

\begin{figure}[tbp]
\centering{\includegraphics[scale=0.25]{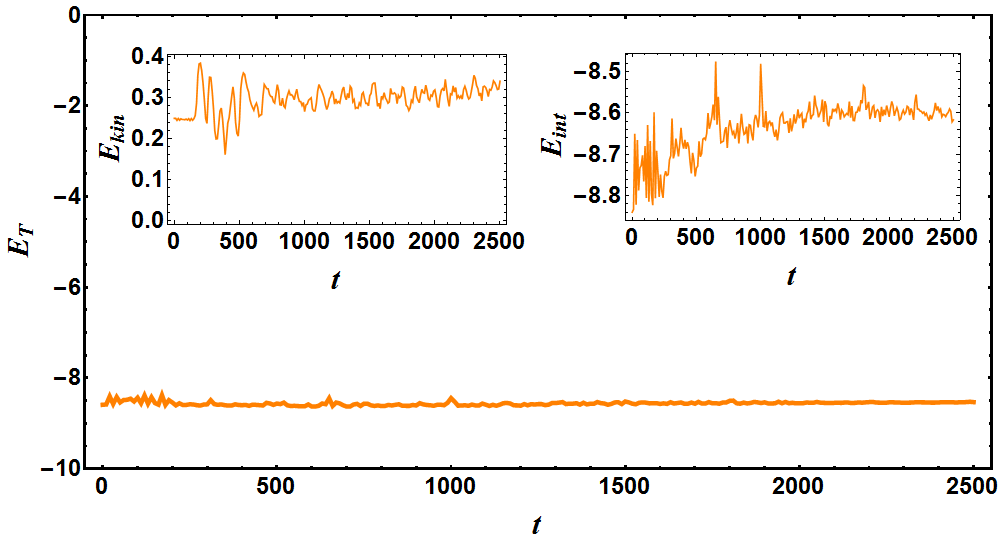}} %
\caption{(Color Online) Variation of the total energy over time. The total energy includes the contribution of the reflected, transmitted and trapped wave. The left inset describes the kinetic energy contribution while the right inset depicts the interaction energy evolution. It is clear from the picture that the total energy is heavily dominated by the interaction energy. The figure actually complements Fig.~\ref{3_part} in the context of energy calculation.}
\label{total_energy}
\end{figure}

To consider the role of energy in the trapping dynamics driven by the
attractive potential, we use Eq. (\ref{H}) for the system's Hamiltonian, and
address the case when the incident QD splits in three fragments, as shown in
Figs. \ref{3_part} (recall it corresponds to the collision of the QD with
the potential well (\ref{barrier}) of depth $V_{0}=0.3$. The results help to
understand the situation in the general case as well.

\begin{figure}[tbp]
\centering{\includegraphics[scale=0.3]{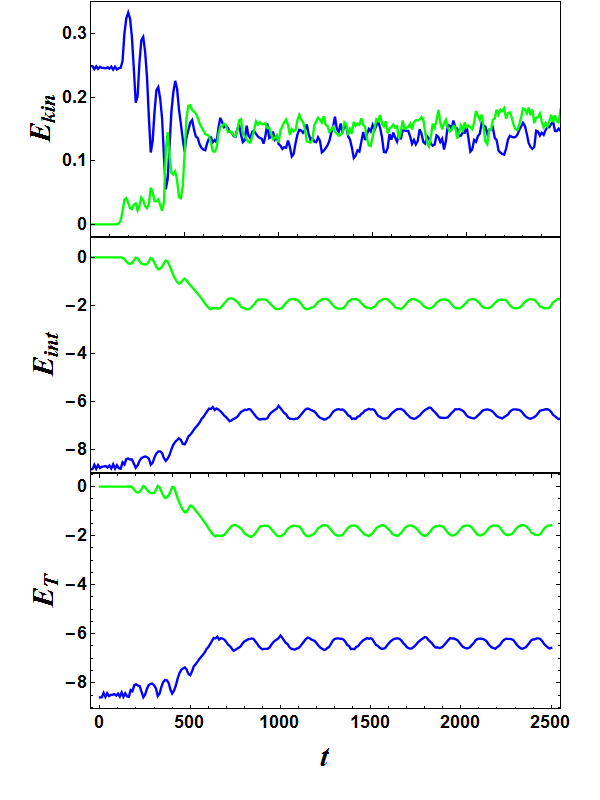}}
\caption{(Color Online) The time evolution of the kinetic energy (top panel), interaction energy (middle panel) and total energy (bottom panel) for the reflected and transmitted components of the soliton. 
The green and blue lines describes the transmitted and reflected waves, respectively.}
\label{part_energy}
\end{figure}

In Fig. \ref{total_energy} variations of the kinetic and interaction
energies are plotted for this case. The plots naturally demonstrate, for the
flat-top QD, the domination of the negative potential energy, which
saturates around a value $-8.6$, over the kinetic energy, which takes values
around $0.2$. Further, Fig. \ref{part_energy} demonstrates that the kinetic
energies of the transmitted and reflected wave packets are nearly equal,
while their velocities and masses (norms) are widely different, as seen in
Fig. \ref{3_part}. Simultaneously, the kinetic energy of the trapped packet
oscillates around $0.006$, indicating the presence of low-frequency
excitations in it. On the other hand, the absolute values of the negative
interaction (potential) energy in the reflected wave function is larger by a
factor $\simeq 3$ than in the transmitted one, due to the obvious fact that
the norm of the reflected packet is essentially larger in Fig. \ref{3_part}.

\begin{figure}[tbp]
\centering{\includegraphics[scale=0.3]{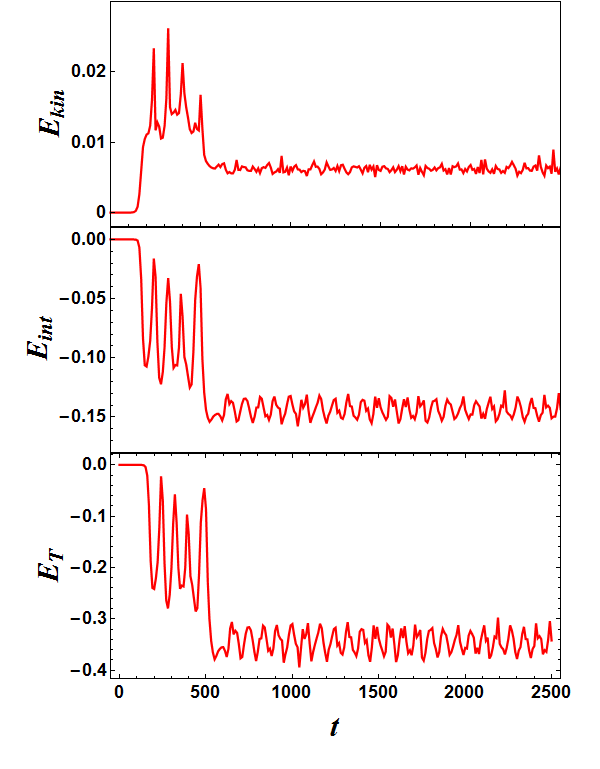}} %
\caption{(Color Online) The figure depicts the time evolution of the kinetic energy (top panel), interaction energy (middle panel) and the total energy (bottom panel) variation of the trapped state created by the
collison of the incident QD with the narrow potential well. The current figure complements Fig.~\ref{3_part} in the context of energy calculation.}
\label{trap_energy}
\end{figure}

Due to the small values of the kinetic and interaction energies in the
trapped wave, they are difficult to discern in Figs. \ref{total_energy} and %
\ref{part_energy}. To make them visible, they are displayed in detail in
Fig. \ref{trap_energy}. As said above, persistent oscillations of the
energies indicate that the trapped wave packet was created in the excited
state. The ratio of the absolute values of the interaction and kinetic
energies for this state is $\simeq 30$, which is comparable to the same
ratios for the reflected and transmitted packets in Figs. \ref{total_energy}
and \ref{part_energy}.

\subsection{Effects of variation of the norm (number of particles) and
velocity on the QD scattering}

\begin{figure}[tbp]
\includegraphics[scale=0.25]{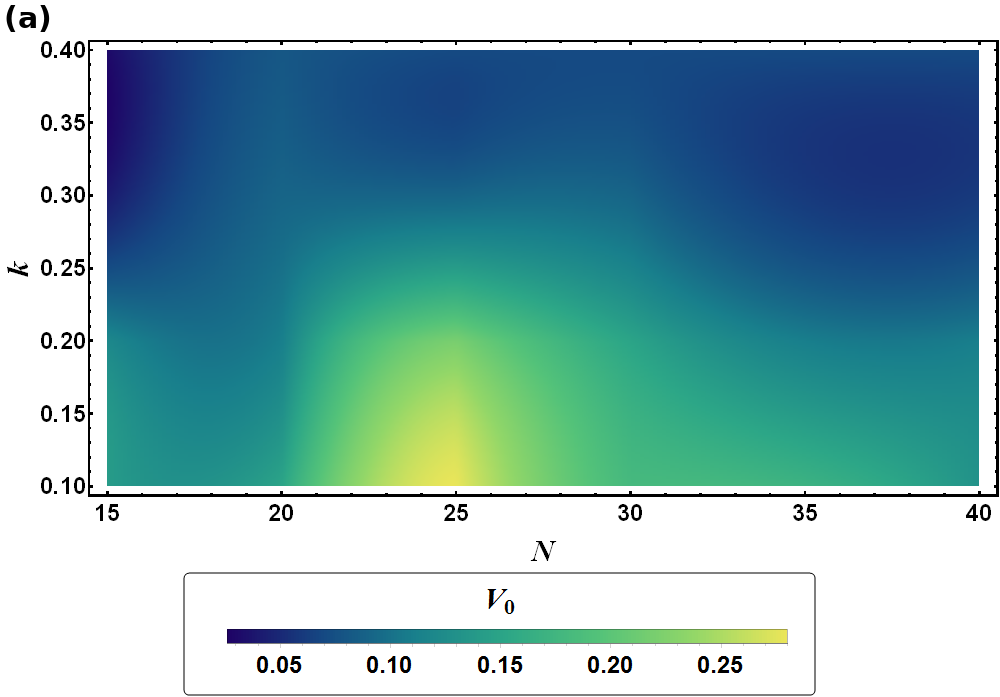} \\
\vspace{0.5cm}
\includegraphics[scale=0.25]{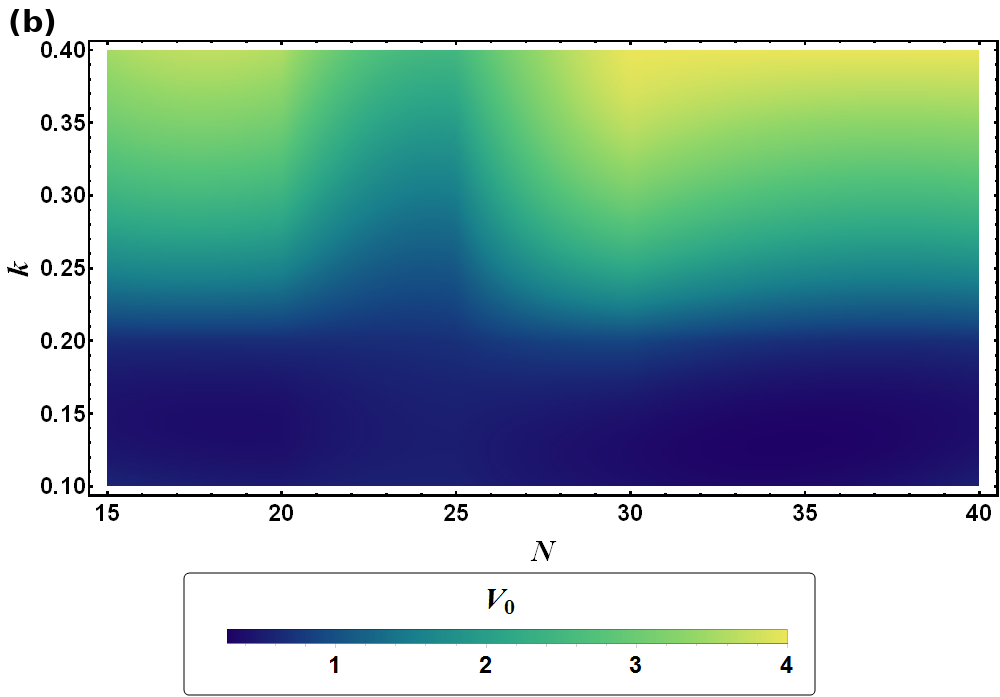}
\caption{(Color Online) (a) The pseudo-color map (\textit{heatmap}) of the maximum value of
the potential-well's depth $V_{0}$ which admits the full transmission (alias
passage) of the incident QD, as a function of the QD's norm $N$ and velocity
$k$. (b) The heatmap of the minimum (threshold) value of $V_{0}$ which
provides the counter-intuitive outcome of the collision, \textit{viz}.,
\emph{full rebound} of the QD from the narrow potential well.}
\label{phase_plot}
\end{figure}

As mentioned earlier, the collision of the QD with a sufficiently deep
potential well tends to make the reflection (rather than the naively
expected transmission) a dominant feature of the interaction [as predicted,
in particular, by Eq. (\ref{effective})]. For this reason, it is interesting
to identify the largest well's depth, $V_{0}$, which still admits the full
transmission (passage) of the incident QD. Figure \ref{phase_plot}(a)
presents a heatmap for such values of $V_{0}$ as a function of the droplet's
norm (scaled number of particles), $N$, and speed, $k$. It is clearly seen
that a maximum value of the so defined depth is achieved for moderate values
of the norm, around $N\simeq 25$. This fact may be understood because the
norm should be relatively large to consider the interaction with the
potential well as a relatively weak perturbation, which cannot produce a
conspicuous effect (such as the reflection) and, on the other hand, if the
norm is too large, it will switch on the TF effect, which gives rise to the
effective repulsive potential, as per Eq. (\ref{effective}). As for the
effect of the velocity $k$, it may be explained, in a coarse approximation,
by noting that, unless the collision is very slow, it excites the
above-mentioned intrinsic oscillations in the trapped wave packet, which
impedes the straightforward transmission.
The above-mentioned counter-intuitive result, namely, the \emph{full rebound}
of the incident QD from the deep narrow potential well, deserves a detailed
consideration too. The result is summarized in the Fig. \ref{phase_plot}(b),
which presents, in the plane of $\left( N,k\right) $, a heatmap of the
minimum (threshold) value of the well's depth, $V_{0}$, above which the full
rebound takes place. The asymptotic independence of the threshold on $N$ is
explained by the fact that the outcome of the collision of the incident
flat-top QD with the narrow potential well is determined by the interaction
of the well with the front (\ref{front}) separating the zero and flat-top
[near-CW, see Eq. (\ref{CW})] levels of the density, irrespective of the
width of the QD pattern.

\begin{figure}[tbp]
\includegraphics[scale=0.25]{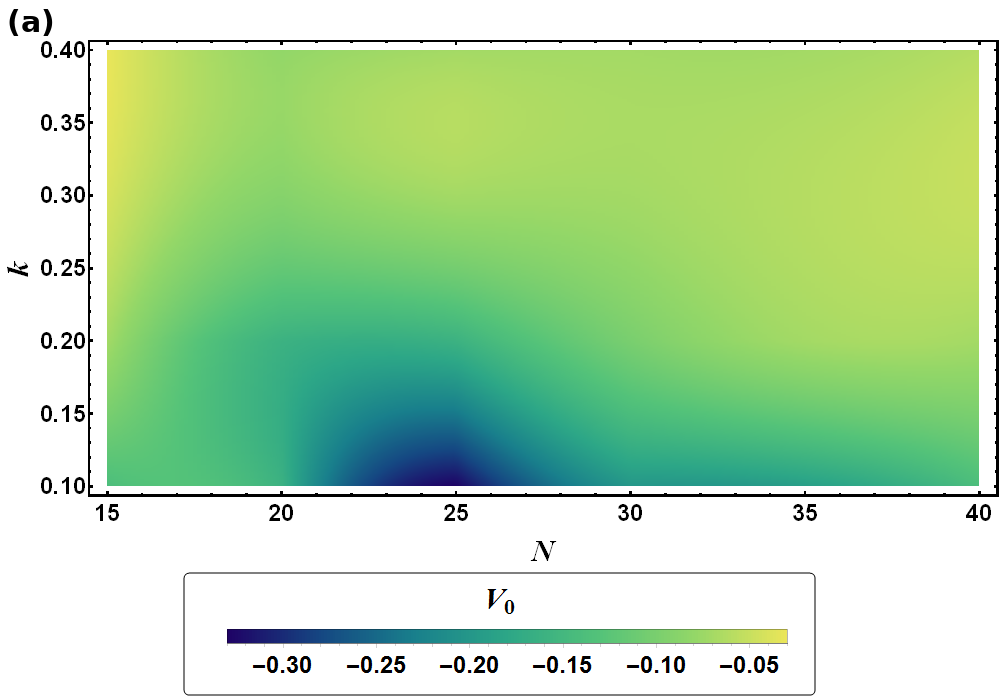} \\
\vspace{0.5cm}
\includegraphics[scale=0.25]{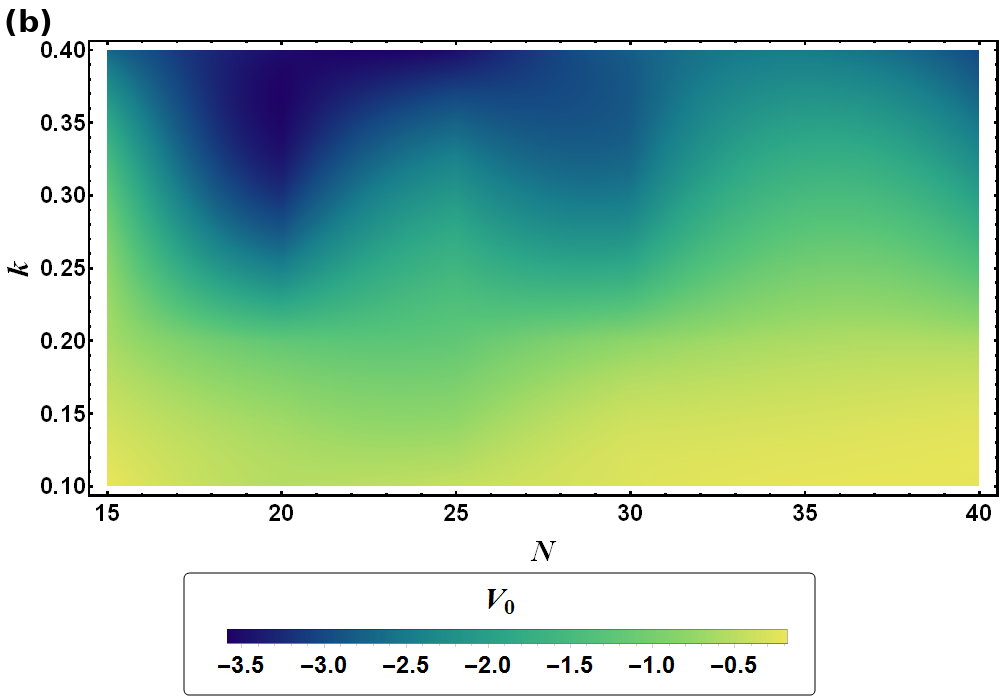}
\caption{(Color Online) (a) The maximum value of the height of the potential barrier, $%
-V_{0}$, which admits full transmission of the incident QD through the
barrier, as a function of $k$ and $N$. (b) The minimum value of $-V_{0}$,
above which full reflection of the incident QD takes place, as a function of
$k$ and $N$.}
\label{phase_wall_plot}
\end{figure}

Finally, we have performed the numerical analysis of the interaction of the
incident QDs with the narrow barrier, which is represented by the narrow
potential (\ref{barrier}) with $V_{0}<0$. The results are summarized in Fig. %
\ref{phase_wall_plot}, in which panel (a) and (b) display, respectively, the
heatmaps, in the $\left( N,k\right) $ plane, for the maximum value of $V_{0}$
which admits the full transmission of the incident QD, and the minimum
(threshold) one above which the total rebound occurs. In the above-mentioned
adiabatic approximation, which neglects deformation of the QD colliding with
the potential barrier \cite{old}, both values of $V_{0}$ would be equal,
given by Eq. (\ref{Uk}) as%
\begin{equation}
V_{0}=9Nk^{2}/8\text{.}  \label{Vlimit}
\end{equation}%
The strong difference between the top and bottom panels of Fig. \ref%
{phase_wall_plot} implies that the deformation, including fission of the
incident QD into the transmitted and reflected fragments, is a conspicuous
effect. Further, it is relevant to mention that, e.g., at $N=20$ and $k=0.25$
Fig. \ref{phase_wall_plot}(b) produces $V_{0}\simeq 2.5$, while Eq. (\ref{Vlimit})
yields, for the same parameters, $V_{0}\simeq 1.406$. The fact that the
numerically found height of the potential barrier necessary for the full
rebound is essentially larger than the prediction of the adiabatic
approximation is explained by the occurrence of the tunneling effect, the
suppression of which requires a taller barrier.

Nevertheless, Eqs. (\ref{thr}) and (\ref{Uk}) make it possible to explain
qualitative features observed in the bottom panel of Fig. \ref%
{phase_wall_plot}. In particular, this is the growth of the threshold value
of $V_{0}$ with the increase of $k$ at fixed $N$.

\section{Conclusion}

\label{conclusion}

Using the known one-dimensional GPE (Gross-Pitaevskii equation) with the
cubic self-repulsion and quadratic attraction induced by quantum
fluctuations, we have studied in detail the static and dynamic solutions for
QDs (quantum droplets) interacting with narrow potential wells and barriers.
In the case of the potential represented by the delta-function potential,
three different stable solutions for QDs pinned to the potential well are
found in the exact form. The TF\ (Thomas-Fermi) approximation for the narrow
rectangular well, and the adiabatic approximation for the collision of the
QD with the potential barrier are elaborated too. Collisions of moving QDs
with the well or barrier are systematically studied by means of direct GPE\
simulations. The results are reported by means of diagrams in the plane of
the QD's norm and velocity, which represent outcomes of the collisions,
including the splitting of the incident QD into transmitted and reflected
fragments, as well as a small trapped one. In particular, the
counter-intuitive outcome in the form of the rebound of the QD from the
potential well is identified and qualitatively explained. In the general
case, the transmitted, reflected and trapped wave packets emerge in excited
states, featuring intrinsic oscillations. Some peculiarities of the
dynamical findings are explained by the analytical predictions.

As mentioned earlier, collisions of matter-wave solitons is a well-studied
topic. Here, we have tried to extend the systematic study for the
interactions QDs with narrow potential wells and barriers. The theoretical
results may be an incentive for experiments with QDs, and also propose new
directions for the development of the matter-wave interferometry.

\section*{Acknowledgment}

The work of B.A.M. was supported, in part, by the Israel Science Foundation
through grant No. 1695/22. AK thanks Department of
Science and Technology (DST), India for the support provided through the project number CRG/2019/000108.

\bibliography{ms_new2}
\end{document}